\documentclass[preprintnumbers,nofootinbib,superscriptaddress,11pt]{revtex4}
\usepackage[hypertex]{hyperref}
\usepackage[dvipdfmx]{graphicx} 
\usepackage{amsmath,amssymb,amsfonts,bm,cancel} 

    \def\d{\delta} \def\D{\Delta} \def\e{\epsilon}   \def\th{\theta}      \def\n{\nu}               

\def\dg{\dagger}  

    \newcommand{\GeV}{ {\rm GeV} }

\newcommand{\lsp}{ \left ( } \newcommand{\rsp}{ \right ) }

\def\abs#1{\left| #1\right|}

\newcommand{\row}[2]{ \begin{pmatrix}  #1 & #2   \end{pmatrix}  }
\newcommand{\column}[2]{ \begin{pmatrix}  #1 \\ #2 \\  \end{pmatrix} }

\newcommand{\Column}[3]{ \begin{pmatrix} #1 \\ #2 \\ #3 \end{pmatrix} }
\newcommand{\diag}[2]{ \begin{pmatrix}  #1 & 0 \\ 0 & #2 \\   \end{pmatrix}  }

\newcommand{\Diag}[3]{ \begin{pmatrix} #1 & 0 & 0 \\ 0 & #2 & 0 \\ 0 & 0 & #3 \\\end{pmatrix}}

\begin{document}

%%%%%%%%%%%%%%%%%%%%%%%%%%%%%%%%%%%%%%%%

\title{\Large $1/(M_{R})_{33}$ expansion of the type-I seesaw mechanism \\
and partial $Z_{2}$ symmetry for TM$_{1,2}$ mixing}

\preprint{STUPP-22-257}
%%%%%%%%%%%%%%%%%%%%%%%%%%%%%%%%%%%%%%%%
\author{Masaki J. S. Yang}
\email{yang@krishna.th.phy.saitama-u.ac.jp}
\affiliation{Department of Physics, Saitama University, 
Shimo-okubo, Sakura-ku, Saitama, 338-8570, Japan}
\affiliation{Department of Physics, Graduate School of Engineering Science,
Yokohama National University, Yokohama, 240-8501, Japan}

%%%%%%%%%%%%%%%%%%%%%%%%%%%%%%%%%%%%%%%%

%\date{\today}

%%%%%%%%%%%%%%%%%%%%%%%%%%%%%
\begin{abstract} %%%%%%%%%%%%%%%%%%%%%
%%%%%%%%%%%%%%%%%%%%%%%%%%%%%

We consider an expansion of the type-I seesaw mechanism by the inverse of the 3-3  matrix element $1/(M_{R})_{33}$ of the mass matrix of right-handed neutrinos $M_{R}$.
Conditions of such a situation are obtained for $M_{R}$ and the Dirac mass matrix $m_{D}$. 

In this case,  a partial $Z_{2}$ symmetry %for the deformed Dirac mass matrix $\tilde m_{D}$ 
such as $S  m_{D} P_{} = \pm  m_{D} P_{}$ with a projection matrix $P_{} = {\rm diag} ( 1, 1, 0)$ leads to an approximate $Z_{2}$ symmetry by $S$ for the neutrino mass matrix $m_{\nu}$. 
Such a partial $Z_{2}$ symmetry is desirable in the context of unified theories 
because it allows hierarchical $m_{D}$ and the large mixing of $m_{\n}$ simultaneously.

%%%%%%%%%%%%%%%%%%%%%%%%%%%%%
\end{abstract} %%%%%%%%%%%%%%%%%%%%%%
%%%%%%%%%%%%%%%%%%%%%%%%%%%%%

\maketitle

%%%%%%%%%%%%%%%
\section{Introduction}
%%%%%%%%%%%%%%%

The minimal seesaw model 
\cite{Ma:1998zg, Frampton:2002qc, Xing:2020ald, Guo:2003cc,Barger:2003gt, Mei:2003gn, Chang:2004wy, Guo:2006qa, Kitabayashi:2007bs, He:2009pt, Ge:2010js, Yang:2011fh, He:2011kn, Harigaya:2012bw, Kitabayashi:2016zec, Bambhaniya:2016rbb, Li:2017zmk, Liu:2017frs, %Shimizu:2017fgu, Shimizu:2017vwi, 
Nath:2018hjx, Barreiros:2018bju, Nath:2018xih,  Wang:2019ovr, Zhao:2020bzx} 
has been studied extensively as a toy model of the three-generation seesaw mechanism 
\cite{Minkowski:1977sc,GellMann:1980v,Yanagida:1979as, Mohapatra:1979ia},
because it is a limit where the mass of the heaviest right-handed neutrino $M_{3}$ is taken to infinity in the three-generation model. 
Decoupling effect of contribution from $M_{3}$ is discussed in the sequential dominance \cite{King:1998jw, King:1999cm, King:1999mb, King:2002nf, Antusch:2004gf, Antusch:2004re,  Antusch:2006cw, Antusch:2010tf}
and in Ref.~\cite{Haba:2008dp}.
In this letter, to consider a situation where $M_{3}$ is large but its contribution is finite, 
the type-I seesaw mechanism in a general basis is expanded by the inverse of the largest 3-3 matrix element $1/(M_{R})_{33}$ of the mass matrix of the right-handed neutrinos $M_{R}$.
If the terms proportional to $1/(M_{R})_{33}$ are sufficiently small, the type-I seesaw mechanism can be described as the sum of the minimal seesaw model and its perturbations.
We investigate its phenomenological consequences.

%%%%%%%%%%%%%%%%%%%%%%
\section{$1/(M_{R})_{33}$ expansion}%
%%%%%%%%%%%%%%%%%%%%%%

In the type-I seesaw mechanism, %\cite{Minkowski:1977sc,GellMann:1980v,Yanagida:1979as, Mohapatra:1979ia}, 
the Dirac mass matrix $m_{D}$ and the symmetric Majorana mass matrix $M_{R}$ of the right-handed neutrinos $\n_{Ri}$ are defined as
\begin{align}
m_{D} =
\begin{pmatrix}
A_1 & B_1 & C_1 \\
A_2 & B_2 & C_2 \\
A_3 & B_3 & C_3 \\
\end{pmatrix} 
\equiv (\bm A \, , \bm B \, , \bm C) \, , 
~~~
M_{R} =
\begin{pmatrix}
 M_{11} & M_{12} & M_{13} \\
 M_{12} & M_{22} & M_{23} \\
 M_{13} & M_{23} & M_{33} \\
\end{pmatrix} \, .
\label{MR}
\end{align}
These matrix elements $A_{i}, B_{i}, C_{i}$ and $M_{ij}$ are general complex parameters. 
By assuming that $M_{R}$ is hierarchical ($|M_{33}| \gg |M_{ij}|$), the absolute value of $M_{33}$ is close to the heaviest mass value $M_{3} \simeq |M_{33}|$.
Since this model is reduced to the minimal seesaw model \cite{Ma:1998zg, King:1998jw, Frampton:2002qc} in the limit of $|M_{33}| \to \infty$,
we consider isolating its contribution.

First, $M_{R}$ is divided into the lighter (first and second) and third generations as, 
\begin{align}
M_{R} \equiv
\begin{pmatrix}
M_{R0} & u \\ 
u^{T} & M_{33}
\end{pmatrix}
\equiv
\lsp \begin{array}{cc|c}
 M_{11} & M_{12} & M_{13} \\
 M_{12} & M_{22} & M_{23} \\ \hline
 M_{13} & M_{23} & M_{33} \\
\end{array} \rsp \, .
\end{align} 
Considering blockwise inversion for $M_{R}$, we obtain a seesaw-like expression; 
\begin{align}
%M_{R}^{-1} = 
 %
\begin{pmatrix}
M_{R0} & u \\ 
u^{T} & M_{33}
\end{pmatrix}^{-1} = 
\begin{pmatrix}
M_{R0}^{-1}+ M_{R0}^{-1} u (M_{R}^{-1})_{33} u^{T} M_{R0}^{-1} &- M_{R0}^{-1} u (M_{R}^{-1})_{33} \\
 - (M_{R}^{-1})_{33} u^{T} M_{R0}^{-1}& (M_{R}^{-1})_{33}
\end{pmatrix} \, ,
\label{3}
\end{align}
where 
\begin{align}
(M_{R}^{-1})_{33} = (M_{33} - u^{T} M_{R0}^{-1} u)^{-1} = { \det M_{R0} \over \det M_{R}} 
\equiv {M_{11} M_{22} - M_{12}^{2} \over (M_{11} M_{22} - M_{12}^{2}) M_{33}  - N } \, , 
\end{align}
with $N = M_{22} M_{13}^2-2 M_{12} M_{23} M_{13}+M_{11} M_{23}^2$. 
The inverse of $M_{R}$ is then
\begin{align}
M_{R}^{-1} = 
\begin{pmatrix}
M_{R0}^{-1} & 0_{2} \\
0_{2}^{T} & 0
\end{pmatrix} 
+ (M_{R}^{-1})_{33}
V \otimes V^{T} \, , 
%=
%\begin{pmatrix}
%X^{-1} Y  \otimes Y^{T} X^{-1}& - X^{-1} Y \\
% -  Y^{T} X^{-1}& 1
%\end{pmatrix} \, ,
\label{5}
\end{align}
where 
\begin{align}
V \equiv \column{- M_{R0}^{-1} u}{1}
 = \Column{(M_{R}^{-1})_{13}  \over (M_{R}^{-1})_{33} }{ (M_{R}^{-1})_{23} \over (M_{R}^{-1})_{33}}{1} 
  = \Column{M_{12} M_{23} - M_{13} M_{22} \over M_{11} M_{22} - M_{12}^{2} }
  { M_{12} M_{13} - M_{11} M_{23} \over M_{11} M_{22} - M_{12}^{2}}{1} \, . 
  \label{6}
\end{align}
Since $M_{33}$ only appears in $\det M_{R}$, 
it is easy to perform a series expansion by $1/|M_{33}| \ll 1/|M_{ij}|$;
\begin{align}
M_{R}^{-1} = 
\begin{pmatrix}
M_{R0}^{-1} & 0 \\
0 & 0
\end{pmatrix} 
+ { 1 \over M_{33}} 
\lsp 1 + \lsp { N \over M_{33} \det M_{R0}} \rsp + \lsp { N \over M_{33} \det M_{R0}} \rsp^{2} + \cdots \rsp
V \otimes V^{T} \, . 
\label{expand}
\end{align}
%
%Incidentally, performing the similar operation again for $M_{R0}^{-1}$ and $M_{22}$ results in a generalized Cholesky decomposition \cite{Yang:2021arl, Yang:2022wch} of the symmetric matrix $M_{R}$.
%A formula for the seesaw mechanism by such a decomposition will be discussed later.

From Eqs.~(\ref{5}) and (\ref{6}), the type-I seesaw mechanism can be written as follows; 
\begin{align}
m_{\n} &= m_{D} M_{R}^{-1} m_{D}^{T} = 
m_{D} \left [ 
\begin{pmatrix}
M_{R0}^{-1} & 0_{2} \\
0_{2}^{T} & 0
\end{pmatrix} 
+ (M_{R}^{-1})_{33}
V \otimes V^{T} 
\right ]
m_{D}^{T} \label{12} \\
%%%
& = m_{D0} M_{R0}^{-1} m_{D0}^{T}
+ 
 (M_{R}^{-1})_{33} \, \bm c \otimes \bm c^{T} \equiv m_{\n 0} + \d m_{\n} \, ,  
\label{13}
\end{align}
where %$m_{D} $ and 
\begin{align}
m_{D0} \equiv  
\begin{pmatrix}
 &  \\[-8pt] 
{\bm A} \, & \bm B  \\[-8pt] 
  & 
\end{pmatrix} , ~~~
\bm c \equiv m_{D} V = %m_{D} \column{- M_{R0}^{-1} u}{1} = 
{ (M_{R}^{-1})_{13} \over  (M_{R}^{-1})_{33} } \bm A +
 {(M_{R}^{-1})_{23} \over (M_{R}^{-1})_{33}} \bm B + \bm C \, . 
 \label{14}
\end{align}
This expression of $m_{\n}$ is valid in any basis (and even for non-hierarchical $M_{R}$) and separates minimal seesaw contributions from the rest. 
Note that factors $(M_{R}^{-1})_{(1,2)3} /  (M_{R}^{-1})_{33}$ are finite in the limit of $M_{33} \to \infty$. 

Moreover, performing the $LDL^{T}$ (or generalized Cholesky) decomposition for the minimal seesaw model, we obtain
\begin{align}
M_{R0}^{-1} &=  {1\over \det M_{R0}} 
\begin{pmatrix}
M_{22} & - M_{12} \\ - M_{12} & M_{11}
\end{pmatrix}
  = 
{1\over \det M_{R0}} 
\begin{pmatrix}
M_{22} & - M_{12} \\ - M_{12} & {M_{12}^{2} \over M_{22}}
\end{pmatrix}
+ \diag{0}{ 1/ M_{22}}  \\
& = 
\begin{pmatrix}
 1 & 0 \\
- { M_{12} \over M_{22}} & 1 \\
\end{pmatrix}
\diag{M_{22} \over \det M_{R0}}{1 \over M_{22}}
\begin{pmatrix}
 1 & -{ M_{12}  \over M_{22}} \\
0 & 1 \\
\end{pmatrix}
\equiv LDL^{T} \, , 
\end{align}
and a formula of $m_{\n 0}$ in a general basis \cite{Yang:2021arl, Yang:2022wch}; 
\begin{align}
m_{\n 0} & = 
{ M_{22} \over \det M_{R0}}
{\bm a} \otimes {\bm a}^{T}
+ {1\over M_{22}}
\bm B \otimes \bm B^{T} \, , ~~~ \bm a = \bm A - \bm B {M_{12} \over M_{22}} \, . 
\label{minimal}
\end{align}
This formula is used in a partial $Z_{2}$ symmetry that will be shown later.

%%%%%%%%%%%%%%%%%%%%%%%%%%%%%%%%
\subsection{Conditions for $\d m_{\n}$ to be considered perturbations}
%%%%%%%%%%%%%%%%%%%%%%%%%%%%%%%%

In Eqs.~(\ref{12}) and (\ref{13}), let us consider a situation where the absolute values of the second term $\d m_{\n}$ are sufficiently smaller than that of $m_{\n 0}$. 
To this end, matrix elements $(\d m_{\n}')_{ij}$ in the diagonalized basis of $m_{\n 0}$ must be regarded as perturbations compared to the singular values of $m_{\n 0}$. 
Probably such conditions will be solutions to complex equations.
Here, to investigate simpler conditions, we focus on the 1-2 block matrix of Eq.~(\ref{3}); 
\begin{align}
(M_{R}^{-1})_{ab} = 
( M_{R0}^{-1} + M_{R0}^{-1} u  (M_{R}^{-1})_{33} u^{T} M_{R0}^{-1})_{ab} \, , 
\label{11}
\end{align}
with $a,b = 1,2$.  
These terms yield $m_{\n0}$ and a part of $\d m_{\n}$, respectively; 
\begin{align}
m_{\n 0} = m_{D0} M_{R0}^{-1} m_{D0}^{T}\, , ~~~
\d m_{\n 0 } \equiv m_{D0} (M_{R0}^{-1} u  (M_{R}^{-1})_{33} u^{T} M_{R0}^{-1}) m_{D0}^{T} \, .
\end{align}
Since $m_{D0}$ is common in $m_{\n 0}$ and $\d m_{\n 0}$, we can consider a  condition by comparing the two terms $M_{R0}^{-1}$ and $M_{R0}^{-1} u (M_{R}^{-1})_{33} u^{T} M_{R0}^{-1}$.
%
%The condition that matrix elements of the second term in Eq.~(\ref{11}) are perturbatively small 
For each matrix element, it is roughly given by
\begin{align}
0.1 |( M_{R0}^{-1})_{ab}| \gtrsim |(M_{R0}^{-1} u  (M_{R}^{-1})_{33} u^{T} M_{R0}^{-1})_{ab}| \, .
\end{align}
From Eq.~(\ref{6}) and $(M_{R}^{-1})_{33} \simeq M_{33}^{-1}$, the condition becomes
\begin{align}
\abs{{0.1 \over M_{33} \det M_{R0} }
\begin{pmatrix}
 M_{22} & -M_{12}  \\
- M_{12} & M_{11} \\
\end{pmatrix}}
\gtrsim 
\abs{\column{(M_{R}^{-1})_{13} }{ (M_{R}^{-1})_{23} } 
\otimes 
\row{(M_{R}^{-1})_{13} }{ (M_{R}^{-1})_{23} } } \, .
\label{cond11}
\end{align}
After all, this inequality is equivalent to comparing the matrix elements of $M_{R0}^{-1}$ with the coefficients of $\bm A, \bm B$ in Eq.~(\ref{14}). 
%この不等式があれば、任意の$\bm A$と$\bm B$について$0.1 |m_{\n0}| \lesssim |\d m_{\n 0}|$が成り立つ。
Given this inequality, $0.1 |m_{\n0}| \gtrsim |\d m_{\n 0}|$ holds for any $\bm A$ and $\bm B$. 
However, this condition does not make sense if there is a zero texture such as $M_{11} = 0$.

A more concise condition is obtained by comparing the matrix elements in the diagonalized basis of $M_{R0}^{-1}$. 
At first we define $D = {\rm diag} (M_{1}^{0}\, , M_{2}^{0})$ where 
 $M_{1,2}^{0}$ is the mass singular values restricted to the minimal seesaw mechanism.
%$ D^{-1} = U^{*} M_{R0}^{-1} U^{\dg} = {\rm diag} (M_{1}^{0} \, , M_{2}^{0})$ 
%
By substituting $M_{R0}^{-1} = U^{T} D^{-1} U$ in Eq.~(\ref{11}) and 
multiplying $D \, U^{*}$ and $U^{\dg} D$  from the left and right, the 1-2 block matrix becomes 
\begin{align}
 U^{T} D^{-1} U & + U^{T} D^{-1} U u  (M_{R}^{-1})_{33} u^{T} U^{T} D^{-1} U \, , \\
D &+ U u  (M_{R}^{-1})_{33} u^{T} U^{T}\, . 
%%%
%0.1 \diag{\det M_{R0} \over M_{22}}{M_{22}} M_{33}  &\gtrsim
%\column {  M_{22}M_{13} - M_{12} M_{23} \over M_{22}}{M_{23}}
%\row {  M_{22}M_{13} - M_{12} M_{23} \over M_{22}}{M_{23}}
\end{align}
By comparing the diagonal elements, 
another condition for $\d m_{\n 0}$ is obtained as 
\begin{align}
0.1 \column{M_{1}^{0}}{M_{2}^{0}} & \gtrsim {1\over M_{33}} \column{ |(U u)_{1}|^{2}}{|(U u)_{2}|^{2}} \, . 
\end{align}
This can be regarded as a perturbative condition in a basis where $M_{R0}^{-1}$ is diagonal. 
Given the hierarchy $|M_{22}| \gg |M_{12}|, |M_{11}|$, the mass values are  roughly evaluated as  $M_{1}^{0} \simeq \abs{\det M_{R0} / M_{22}}, M_{2}^{0} \simeq |M_{22}| $ and $U \sim 1$ holds.
Thus the perturbative condition is approximately
\begin{align} 
 0.1|M_{33}| \column{ \abs{\det M_{R0} \over M_{22} }  }{ \abs{ M_{22} } } 
\gtrsim \column{|M_{13}|^{2}}{|M_{23}|^{2}} \, . 
\label{cond12}
\end{align}
Eqs.~(\ref{cond11}) and (\ref{cond12}) are inequalities for absolute values of the 2-dimensional vectors $M_{R0}^{-1} u = U^{T} D^{-1} U u$ and $Uu$.
Thus,  if $U$ is small mixing $U \sim 1$, 
the two inequalities are approximately equivalent through the scale transformation by $D$. 

In the limit where the off-diagonal elements $M_{ij}$ are small, 
the upper bound of Eq.~(\ref{cond12}) corresponds to the singular values $M_{1,2,3}$ of $M_{R}$.
Thus, these bounds approximately lead to the following form of $M_{R}$; 
\begin{align}
M_{R} \sim 
\begin{pmatrix}
\e & * & |M_{13}| \lesssim 0.3 \, \sqrt{\e M_{33}}  \\
 & \d & |M_{23} |\lesssim 0.3 \, \sqrt{\d M_{33} }\\
 & & M_{33} \\
\end{pmatrix} \, .
\end{align}
Here, $O(1)$ coefficients are omitted and there is not much restriction on $M_{12}$ as long as $|M_{12} / M_{22}| \lesssim 1$ holds. 

Similarly, focusing on the 3-3 element of Eq.~(\ref{3}) or the term of $\bm C \otimes \bm C^{T}$ in Eq.~(\ref{14}), we obtain  a new condition for $\bm C$; 
\begin{align}
\abs{C_{i}^{2} \over M_{33} }  \lesssim  0.1  \{ \abs{A_{i}^{2} M_{22} \over \det M_{R0} }, \abs{B_{i}^{2} M_{11} \over \det M_{R0}}  \}  \, . 
\label{cond2}
\end{align}
In other words, the condition is simply written as
$|C_{i}|^{2} \lesssim 0.1 |M_{33}| \{ \sqrt{\D m_{21}^{2}} \, , \sqrt{ \D m_{31}^{2}} \}$ by the  mass squared differences $\D m_{ij}^{2}$. 
If the magnitude of $C_{3} = (m_{D})_{33}$
% = {v\over \sqrt 2} (Y_{\n})_{33}$ (with the vacuum expectation value of the Higgs field $v$) 
is close to the mass of the top quark $m_{t}$, we obtain a lower bound for $M_{3}$; 
\begin{align}
 M_{3} \gtrsim {10 \, m_{t}^{2} \over \sqrt{\D m_{31}^{2}} } \simeq 5.81 \times 10^{15} \, \GeV \, . 
 % \D m_{21}^{2} = 74.2 meV^{2}
\end{align}
Although this fact has been mentioned in Ref.~\cite{Haba:2008dp}, 
interestingly, the lower bound is quite close to the scale of grand unified theories (GUTs).
When these two perturbative conditions (Eq.~(\ref{cond11}) or (\ref{cond12}) and Eq.~(\ref{cond2})) is satisfied, the contribution from the off-diagonal block in Eq.~(\ref{3}) (or cross terms such as $\bm A \otimes \bm C^{T}$ and $\bm B \otimes \bm C^{T}$ in Eq.~(\ref{14})) is automatically small. Therefore two conditions are sufficient.

%%%%%%%%%%%%%%%%%%%%%%%%
\subsection{Partial $Z_{2}$ symmetry }
%%%%%%%%%%%%%%%%%%%%%%%%

In such a ``quasi-minimal'' seesaw model, a partial $Z_{2}$ symmetry of $m_{D}$ leads to an approximate $Z_{2}$ symmetry of $m_{\n}$.
The following TM$_{1,2}$ mixing \cite{Albright:2008rp, Albright:2010ap, He:2011gb} is well discussed in many seesaw models \cite{ Luhn:2013lkn, Li:2013jya, Shimizu:2017fgu, Shimizu:2017vwi,  King:2019vhv, Krishnan:2020xeq};
\begin{align}
U_{\rm TM1} = U_{\rm TBM} U_{23} \, , ~~~
U_{\rm TM2} = U_{\rm TBM} U_{13} \, , ~~~
\end{align}
where 
\begin{align}
U_{\rm TBM} 
= 
\begin{pmatrix}
 \sqrt{\frac{2}{3}} & \frac{1}{\sqrt{3}} & 0 \\
 -\frac{1}{\sqrt{6}} &\frac{1}{\sqrt{3}} & \frac{1}{\sqrt{2}} \\
 -\frac{1}{\sqrt{6}} & \frac{1}{\sqrt{3}} & -\frac{1}{\sqrt{2}} \\
\end{pmatrix} , 
~~~
U_{23} = 
\begin{pmatrix}
1 & 0 & 0 \\
0 & c_{23} & s_{23} e^{- i\phi} \\
0 & - s_{23} e^{i \phi} & c_{23}
\end{pmatrix} ,
~~~ 
U_{13} =
\begin{pmatrix}
c_{13} & 0 & s_{13} e^{- i \phi} \\
0 & 1 & 0 \\
- s_{13} e^{i \phi} & 0 & c_{13}
\end{pmatrix} \, , 
\end{align}
with $c_{ij} \equiv \cos \th_{ij} , \, s_{ij} \equiv \sin \th_{ij}$. 
The mass matrix $m_{\n}$ that predicts TM$_{1,2}$ mixing has a $Z_{2}$ symmetry
$S_{1,2} \, m_{\n} \, S_{1,2} = m_{\n}$ with \cite{Lam:2006wy, Lam:2006wm,Lam:2007qc,Lam:2008rs}
%by the following $S_{1,2} \equiv 1 - 2 \bm v_{1,2} \otimes \bm v_{1,2}^{T}$; 
%
\begin{align}
S_{1}  = 
{1\over 3}
\begin{pmatrix}
-1 & 2 & 2 \\
2 & 2 & -1 \\
2 & -1 & 2 \\
\end{pmatrix} \, , ~~~ 
S_{2} = 
{1\over 3}
\begin{pmatrix}
1 & -2 & -2 \\
- 2 & 1 & -2 \\
- 2 & -2 & 1 \\
\end{pmatrix} .
\label{S2}
\end{align}

In the model where the contribution of $1/M_{33}$ can be regarded as perturbations, 
the conditions for $Z_{2}$ symmetry will be relaxed.
If $\tilde m_{\n D} \equiv (\bm a \, , \bm B \, , \bm C )$ has the following {\it partial} $Z_{2}$ symmetry,  $m_{\n}$ has the $Z_{2}$ symmetry in a good approximation;
\begin{align}
S_{1,2} \tilde m_{D} P = \pm  \tilde m_{D} \, ~~ {\rm or} ~~ S_{1,2} \tilde m_{D} P'  = \pm \tilde m_{D} \, ,
\end{align}
where $P= {\rm diag} (1,1,0)$ and $P' = {\rm diag} (1, -1,0)$ are projections to the lighter generations.
This is because, from Eq.~(\ref{minimal}), 
\begin{align}
S_{1,2} m_{\n 0} S_{1,2} = S_{1,2} \tilde m_{D} P \Diag{M_{22} \over \det M_{R0}}{1\over M_{22}}{0} P \tilde m_{D}^{T} S_{1,2}  = m_{\n 0} \, , 
\end{align}
and $\d m_{\n}$ can be regarded as a perturbation.
Although the first condition $S_{1,2} \tilde m_{D} P = \pm \tilde m_{D} P$ leads to partial $Z_{2}$-symmetric $m_{D}$ (without tilde),
the case $S_{2} m_{D} P = \pm m_{D} P$ is unsuitable because the limit of $M_{33} \to \infty$ predicts $m_{2} = 0$ or $m_{1,3} = 0$~\cite{Yang:2022yqw}. 
Such a situation is similar to the constrained sequential dominance 
\cite{King:2005bj, Antusch:2007dj, Antusch:2011ic, Antusch:2013wn, Bjorkeroth:2014vha}. 
This partial $Z_{2}$ symmetry can be regarded as its extension 
because it predicts TM$_{1,2}$ for general parameter regions and general basis where $M_{R}$ is not diagonal.
Such a partial $Z_{2}$ symmetry is desirable in the context of GUTs 
because it allows hierarchical Dirac neutrino mass $m_{D}$ and the large mixing of $m_{\n}$ simultaneously.

%%%%%%%%%%%%%%
\section{Summary}
%%%%%%%%%%%%%%

To summarize, in this letter, we consider an expansion of the type-I seesaw mechanism by the inverse of the 3-3 matrix element $1/(M_{R})_{33}$ of the right-handed neutrinos.
If the terms proportional to $1/(M_{R})_{33}$ are sufficiently small, 
the type-I seesaw mechanism can be described as the sum of the minimal seesaw model and its perturbations.
We obtain perturbative conditions of such a situation for the mass matrix of the right-handed neutrinos $M_{R}$ and the Dirac mass matrix $m_{D}$. The conditions are approximately written by 
$| (M_{R})_{13} |^{2} \lesssim 0.1 M_{1} M_{3}$, 
$| (M_{R})_{23} |^{2} \lesssim 0.1 M_{2} M_{3}$ and 
$(m_{D})_{i 3}^{2} \lesssim 0.1 M_{3} \{ \sqrt{\D m_{21}^{2}} , \sqrt{\D m_{31}^{2}} \}$ with the mass singular values $M_{i}$ of $M_{R}$.

Moreover, conditions that the neutrino mass matrix $m_{\n}$ has $Z_{2}$ symmetry is relaxed. 
It is found that $m_{\n}$ has an approximate $Z_{2}$ symmetry  
when the deformed Dirac mass matrix $\tilde m_{D}$  has a partial $Z_{2}$ symmetry 
such as $S \tilde m_{D} P_{} = \pm \tilde m_{D} P_{}$,  
restricted by a projection matrix $P_{} = {\rm diag} ( 1, 1, 0)$ to the two lighter generations.

%%%%%%%%%%%%%%
\section*{Acknowledgment}
%%%%%%%%%%%%%%
This study is financially supported %by JSPS KAKENHI Grants
by JSPS Grants-in-Aid for Scientific Research
No.~JP18H01210 and MEXT KAKENHI Grant No.~JP18H05543.

%\bibliographystyle{bib/h-physrev50}
%\bibliography{bib/fourzero,bib/mutausym,bib/trimaximal,bib/refsym,bib/GCP,bib/LR,bib/PSGUT, bib/SD,bib/minimal-natural, bib/TM12}

\end{document}